\documentclass[usenatbib]{an}
\usepackage{natbib}
\usepackage{graphicx}
\usepackage{times}
\usepackage{fancyhdr}
\newcommand{\muG}{\mu{\rm G}}

\begin{document}

\title{Simulating large-scale structure formation with magnetic fields}

\author{Klaus Dolag}
\institute{Max-Planck-Institut f\"ur Astrophysik, 
P.O. Box 1317, D-85741 Garching, Germany}

\date{Received; accepted; published online}

\abstract{
In the past, different works based on numerical simulations have been
presented to explain magnetic fields (MFs) in the large scale structure and
within galaxy clusters.  In this review, I will summarize the main
findings obtained by different authors and - even if many details are
still unclear - I will try to construct a consistent picture of our
interpretation of large-scale magnetic fields based on numerical
effort. I will also sketch how this is related to our understanding of
radio emission and summarize some arguments where our theoretical
understanding has to be improved to match the observations.
}
\correspondence{kdolag@mpa-garching.mpg.de}

\maketitle

\section{Introduction}

There are several different approaches to study the build-up of
MFs in the intergalactic medium. Many papers
consider  MHD simulations of cloud-wind interaction (see
\citealt{2000ApJ...543..775G} and references therein) or are
simulating the rise of relic radio bubbles (see
\citealt{2005ApJ...624..586J}, \citealt{2005MNRAS.357..242R} and
references herein). Such work is usually focused more on the relevance
of the local MFs within these processes, whereas in this review I will
concentrate more on the discussion of MHD simulations which aim at
understanding the build-up of the cluster MF and its possible
origin. So far, firm evidence for the presence of extended MFs has
been found only in galaxy clusters. For recent reviews see
\citet{Carilli.Taylor..2002} and \citet{2004IJMPD..13.1549G}. 
Recently there are observational claims of a detection of substantial
MFs also within the large scale structure (LSS), see contribution by 
Kronberg.
% and coauthors. 
The origin of the MF in
galaxy clusters is still under debate. The variety of possible
contributors ranges from primordial fields, battery and dynamo fields
over all classes of astrophysical objects {which contribute with 
their ejecta.  The later possibility is supported by the observation
of the metal enrichment of the intracluster medium (ICM), which has to
be originated as ejecta of astrophysical objects. In addition, the MFs
produced by all these contributors will be compressed and amplified by
the process of structure formation. The exact amount of this
amplification and the resulting MF filling factor will depend on
place and time at which the contributer is thought to be most
efficient.

\section{Possible origins of MFs within the LSS}\label{sec:origin}

In the following I will describe the three main classes of models for
the origin of cosmological MF within the LSS.

In the first one, MFs are assumed to be produced `locally' at
relatively low redshifts ($z \sim 2-3$) by galactic
(e.g. \citealt{Volk&Atoyan..ApJ.2000}) or AGN ejecta
(e.g. \citealt{Furlanetto&Loeb..ApJ2001}). 
One of the main arguments in favor of this model is that the high
metallicity observed in the ICM suggests 
a significant enrichment driven by galactic winds or AGNs 
in the past.
Winds and jets should carry MFs together with the processed matter.
While it has been shown that winds from ordinary galaxies give rise to
MFs which are far weaker than those observed in galaxy clusters, MFs
produced by the ejecta of starburst galaxies can be as large as
$0.1\,\muG$.  Clearly, this class of models predicts that MFs are
mainly concentrated in galaxy clusters. Note that, if the magnetic
pollution happens early enough (around $z \sim 3$), these fields will
not only be amplified by the adiabatic compression of the
proto-cluster region, but also by shear flows, turbulent motions, and
merging events during the formation of galaxy clusters.

In the second class of models, the MF seeds are assumed to be produced
at higher redshifts, before galaxy clusters form gravitationally
bound systems. Although the strength of the seed field is expected to
be considerably smaller than in the previous scenario, the adiabatic
compression of the gas and the shear flows driven by the accretion of
structures can give rise to a considerable amplification of the MFs.
Several mechanisms have been proposed to explain the origin of MF
seeds at high redshift. Some of them are similar to those discussed
above, differing only in the time at which the magnetic pollution is
assumed to take place.  In the present class of models the MF seeds
are supposed to be expelled by an early population of dwarf starburst
galaxies or by AGN at a redshift between 4 and 6
(\citealt*{Kronberg..1999ApJ}), allowing them to magnetize a large
fraction of the volume.  Alternative models invoke processes that took
place in the early universe. Indeed, the ubiquity of MFs in the
universe suggests that they may have a cosmological origin.  In
general, all `high-$z$ models' predict MF seeds filling the entire
volume of the universe. However, the assumed coherence length of the
field crucially depends on the details of the models. While scenarios
based on phase transitions give rise to coherence lengths which are so
small that the corresponding fields have probably been dissipated, MFs
generated at neutrino or photon decoupling have much higher chances to
survive until the present time. Another (speculative) possibility is
that the seed field was produced during inflation. In this case, the
coherence length can be as large as the Hubble radius. See
\citet{Grasso..PhysRep.2000} for a recent review.

\begin{figure}[t]
\begin{center}
\includegraphics[width=0.4\textwidth]{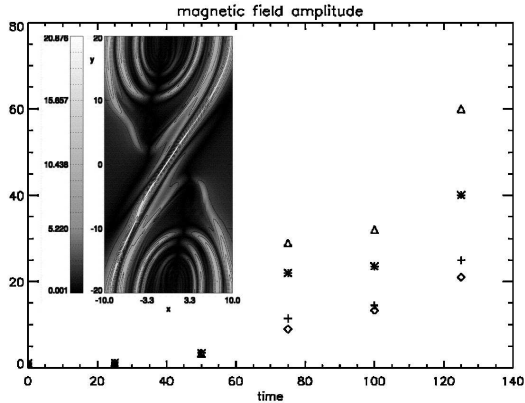}
\end{center}
\caption{{\small Example of MF (y-component)
within an evolving KH instability (inlay) and its time evolution.  
Taken from \citet{1999ApJ...518..177B}.  }}
\label{fig:Birk}
\end{figure}

The third scenario assumes that the MF seeds were produced by the
so-called Biermann battery effect
(\citealt{Kulsrud..ApJ.1997},\citealt*{Ryu..1998}). 
The idea here is that merger shocks related to the hierarchical
structure formation process give rise to small thermionic electric
currents which, in turn, may generate MFs.  The battery process has
the attractive feature of being independent of unknown physics at high
redshift. Its drawback is that, due to the large conductivity of the
IGM, it can give rise to at most very tiny MFs, of order $10^{-21}$
G. One therefore needs to invoke a subsequent turbulent dynamo to
boost the field strength to the observed level. Such a turbulent
amplification, however, cannot be simulated numerically yet, making it
quite difficult to predict how it would proceed in a realistic
environment.  It is clear that one expects the level of turbulence to
be strongly dependent on the environment, and that it should mostly
appear in high-density regions like collapsed objects.  While
energetic events such as mergers of galaxy clusters can be easily
considered to drive the required levels of turbulence, this is harder to
understand in relatively quiet regions like filaments.  Lacking a
theoretical understanding of the turbulent amplification, it is
therefore not straightforward to relate the very weak seed fields
produced by the battery process to the MFs observed today. Attempts to
construct such models based on combining numerical and analytical
computations have not been reported to successful reproduce the
observed scaling relations of MFs in galaxy clusters so far.

\begin{figure}
\includegraphics[width=0.24\textwidth]{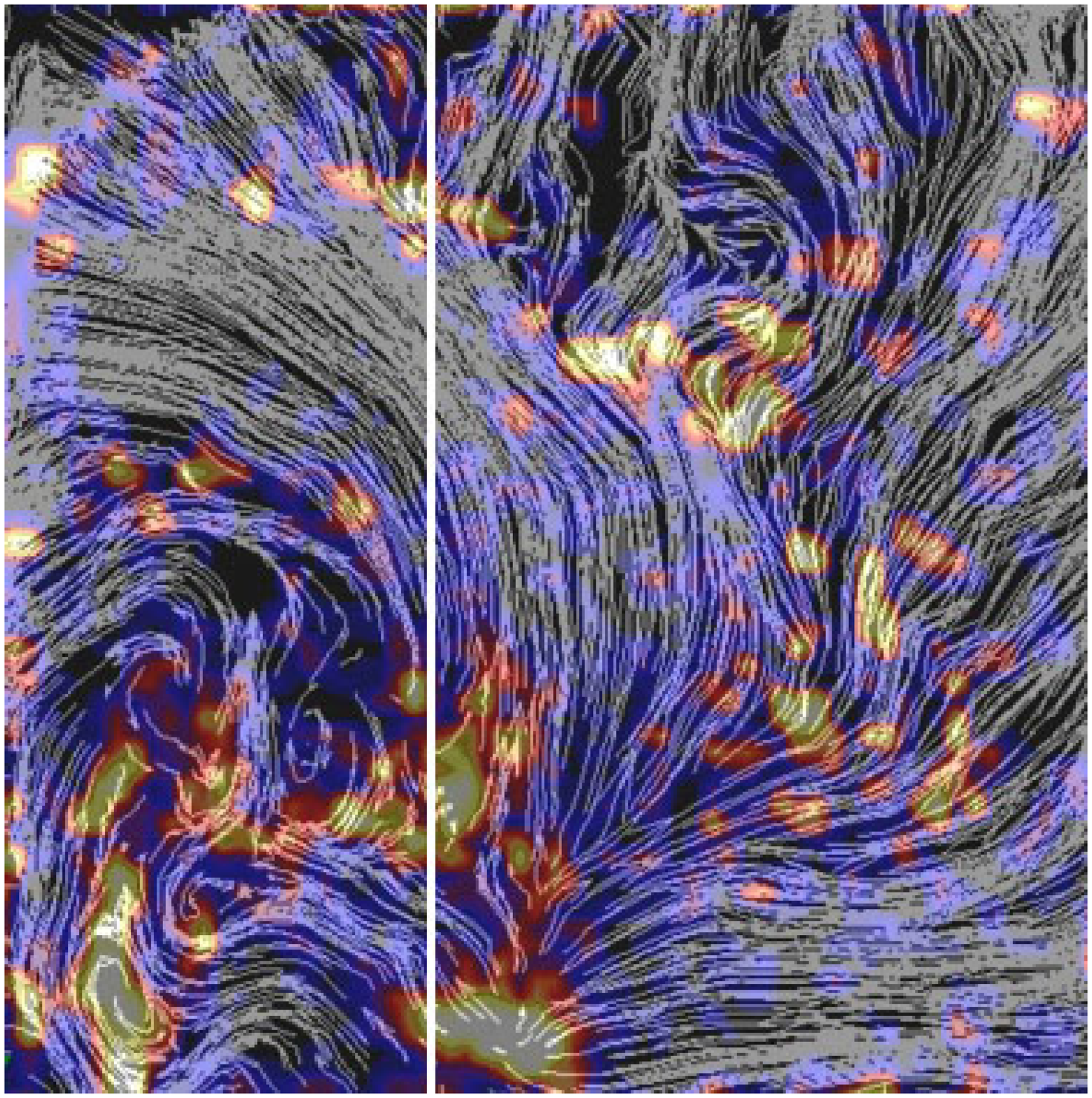}
\includegraphics[width=0.24\textwidth]{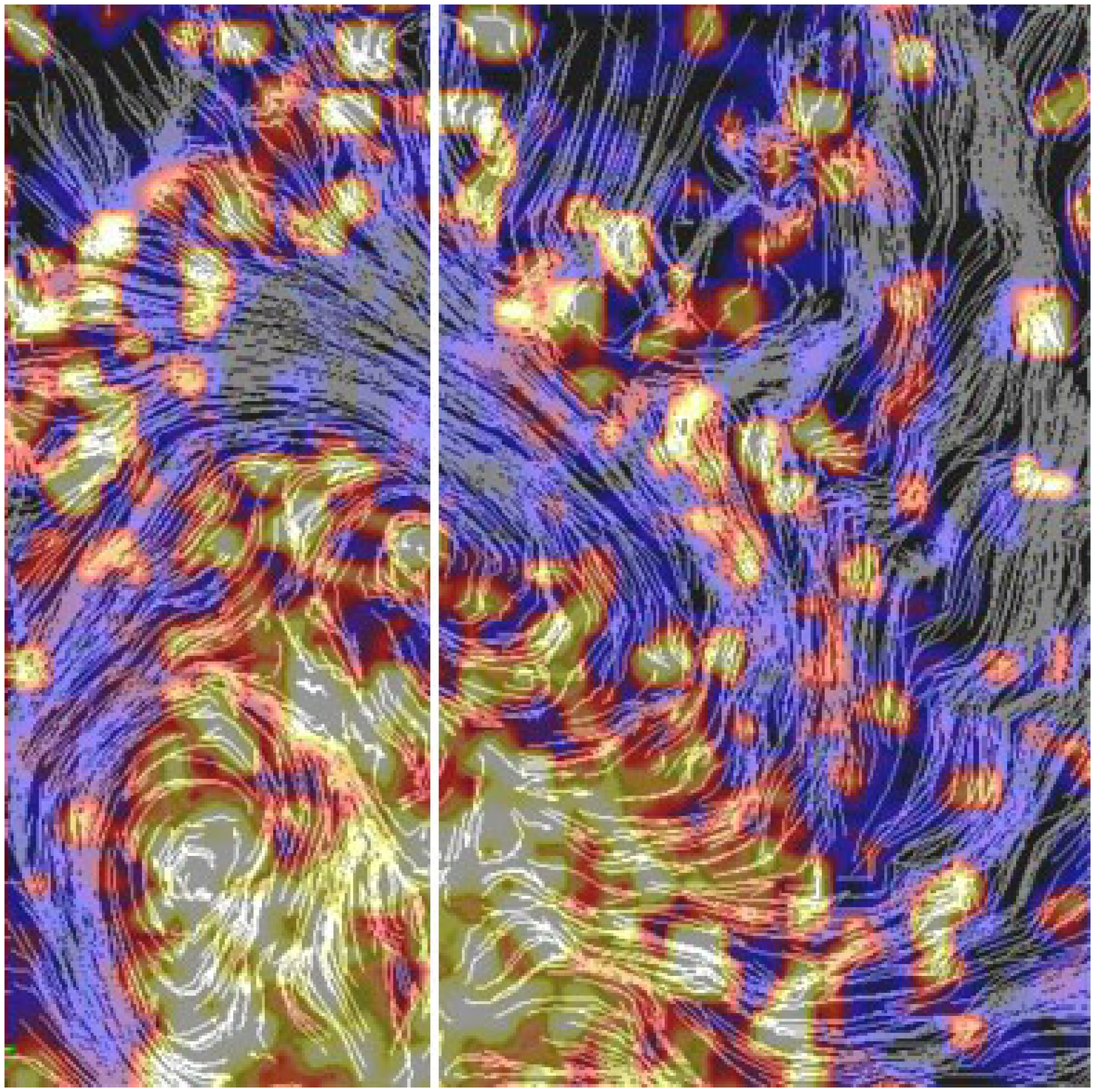}\\
\caption{{\small The gas velocity field in a slice through the central Mpc of
a cluster simulation after subtracting the {\em global} mean bulk
velocity of the cluster. The left panel shows a run with
the original SPH viscosity, the right panel for a low-viscosity
scheme. The underlying color
maps represent the ratio of turbulent kinetic energy and total
kinetic energy content of particles, inferred after substracting
the {\em local} mean velocity, as described in
\citet{2005MNRAS.364..753D}}} \label{fig:dolag_turb}
\end{figure}

\section{Local amplification of MFs}

A very basic process of amplification for MFs is related to the
Kelvin-Helmholtz (KH) instabilities driven by shear flows, which are
common within the dynamics of the structure formation process.
\citet*{1999ApJ...518..177B} performed a detailed study of such
an amplification within the environment of galactic outflows (see
Fig.~\ref{fig:Birk}) in starburst galaxies where the KH timescale
should be $\approx 4\times10^5$ years. Using a Cartesian resistive MHD
code they found that the obtained amplification factor for the MF
mainly depends on the initial ratio of magnetic to kinetic energy and
only mildly depends on the assumed resistivity.  They reached the
conclusion that such a process could indeed explain the significantly
higher MF observed in the starburst galaxy halo compared to what is
expected from the MFs observed within galactic disks. When applied
to a cluster core environment, the KH timescale turns out to be $10^7$
years, making it an interesting process for further amplifying weak
MFs.

\begin{figure}
\includegraphics[width=0.49\textwidth]{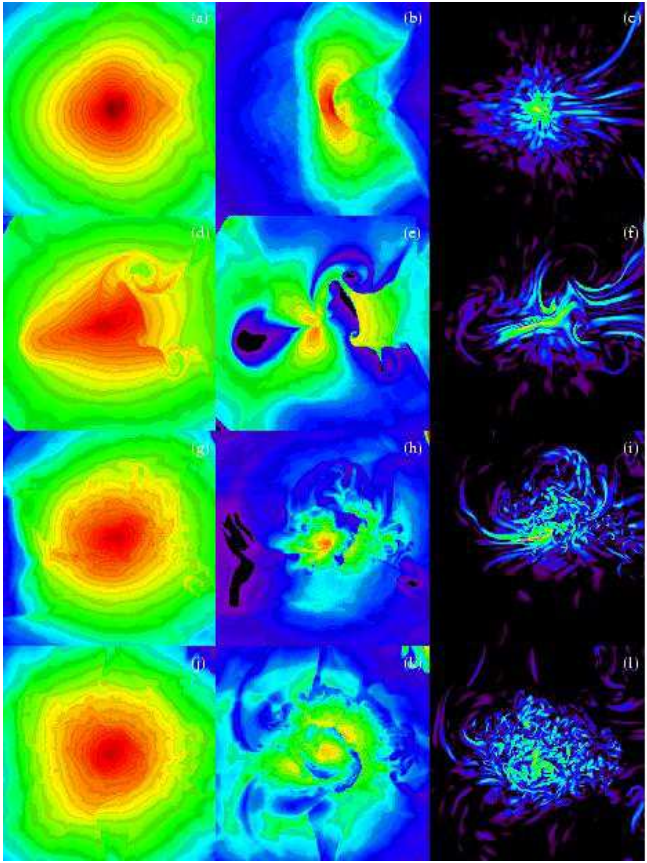}
\caption{{\small The evolution of
(logarithm of) gas density (left column), gas temperature (central
column ), and (logarithm of) magnetic pressure (right column) in
two-dimensional slices taken through the cluster core in the plane of
the merger.  Each row refers to different epochs: $t=0$ (i.e. the time
of core coincidence), $t=1.3$, $t=3.4$, and $t=5.0$ Gyrs, from top to
down.  Each panel is $3.75 \times 3.75$ Mpc. Taken from
\citet{1999ApJ...518..594R}.}}
\label{fig:roet_I_a}
\end{figure}

In recent high-resolution smoothed particle hydrodynamic (SPH) simulations of 
galaxy clusters within a cosmological environment, 
%by adding a novel scheme for treating the 
using a novel scheme to treat 
artificial viscosity within the {\it Gadget2} code
(\citealt{2005MNRAS.364.1105S}), 
%we were able (\citealt{2005MNRAS.364..753D}) to demonstrate 
\citet{2005MNRAS.364..753D} demonstrated
how such shear flows, which are quite common in the
process of cosmic structure formation, drive fluid instabilities and
increase the turbulence level within the ICM to a significant level
(see Fig. \ref{fig:dolag_turb}). It was also shown that the artificial
viscosity used in standard SPH simulations might significantly
suppress such fluid instabilities.

In earlier work, extensive MHD simulations of single merging events
performed using the Eulerian code {\it ZEUS}
(\citealt{1992ApJS...80..753S,1992ApJS...80..791S}) demonstrated the MF
amplification in such merger events (\citealt*{1999ApJ...518..594R}).  In
particular they found that first, the field becomes quite filamentary
as a result of stretching and compression caused by shocks and bulk
flows during infall, but only a minimal amplification occurs. Second,
field amplification is more rapid, particularly in localized regions,
as the bulk flow is replaced by turbulent motions (e.g., eddies), see
Fig \ref{fig:roet_I_a}. The
total MF energy is found to increase by nearly a factor of three with
respect to a non-merging cluster.  In localized regions (associated with
high vorticity), the magnetic energy can increase by a factor of 20 or
more. A power spectrum analysis of the magnetic energy showed that the
amplification is largely confined to scales comparable to or smaller
than the cluster cores: this indicates that the core dimensions define
the injection scale.  It is worth to notice that, due to the lack of
resolution, the previous results can be considered a lower limit on
the total amplification. Furthermore, it is quite likely that a galaxy
cluster undergoes more than one such an event during its formation
process, and that also the accretion of smaller haloes injects
turbulent motions within the ICM: consequently the MF amplification
within galaxy clusters will be even higher.  A detailed discussion of
the amplification of MF in a cluster environment, using various
simulations of driven turbulence, can be found in
\citet*{2005astro.ph..5144S}, where it is 
%demonstrated 
shown
that
reasonable strength and length scales for galaxy clusters can be
obtained by turbulent processes.

\begin{figure}
\includegraphics[width=0.49\textwidth]{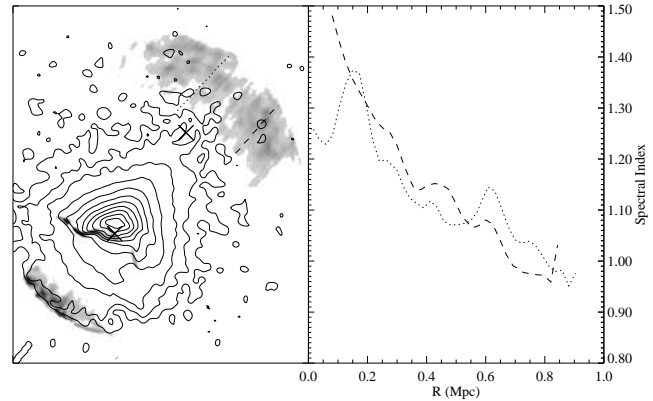}
\caption{{\small In the left panel
we show the simulated A3667 X-ray surface brightness and radio data.
Contours represent the X-ray surface brightness. The grayscale
represents the synthetic radio emission at 1.4 GHz. The image is $3.15
\times 3.85$ Mpc. Dashed and dotted lines refer to the location of the
radio spectral index ($\alpha^{1.4}_{4.9}$) profiles displayed in the
right panel.  Taken from \citet{1999ApJ...518..603R}.}}
\label{fig:roet_II_a}
\end{figure}

\section{The quest for radio relics}

By extending such merger simulations to reproduce the observed X-ray
properties of A3667 and adding a model for in situ re-acceleration of
relativistic particles, \citet*{1999ApJ...518..603R} were able to
reproduce the main features of the extended peripheral radio emission
(the so-called radio relics) observed in A3667. In their models
they injected relativistic electrons with a power-law spectrum, where
the power-law index $\gamma = 3/(r-1)+1$ is related to the gas
compression rate $r$ at the shock.  They also relate the age of the
radio plasma $t_a$ to the distance $d=\kappa v_st_a$, using a
weak-field/high-diffusion limit $\kappa=1$. Having effective shock
velocities $v_s\approx 700-1000 \mathrm{km}/\mathrm{s}$ and aging the
synchrotron spectrum using the formalism by
\citet{1985ApJ...291...52M} they have been able to reproduce the
observed distribution of spectral index for a MF of $\approx 0.6\mu$G
at the position of the radio relic, see Fig. \ref{fig:roet_II_a}. Since such configurations seem to be
quite common in galaxy clusters, naturally the question arises, why
not all clusters show such a peripheral radio emission. One possible
explanation is that such shock structures are relatively short lived
compared to the merger event itself. Also, it is necessary to have the
presence of a large-scale MF. It could further be that only massive
clusters can provide enough MF and strong enough merger events to
trigger such a peripheral emission.

To overcome this problem, \citet{2002MNRAS.331.1011E} proposed that
such radio relics could be made by pre-existing fossil radio plasma
illuminated via shock waves initiated by merger events.  In their work
they evolved the electron spectrum for the tracer particles,
representing the fossil radio plasma, following
\citet{2001ASPC..250..454E} which take into account synchrotron,
inverse Compton and adiabatic energy losses and gains. Their
simulation, using the {\it ZEUS} code, follows the evolution
of a sphere of tracer particles hit by a shock front (see
Fig. \ref{fig:ens_brue_I}).  Such a configuration nicely reproduces
the filamentary radio emission and toroidal structures, as observed in
many cases. These simulations also predict the MFs to be mostly
aligned with the direction of the filaments, as suggested by
observational data.

\begin{figure}
\includegraphics[width=0.49\textwidth]{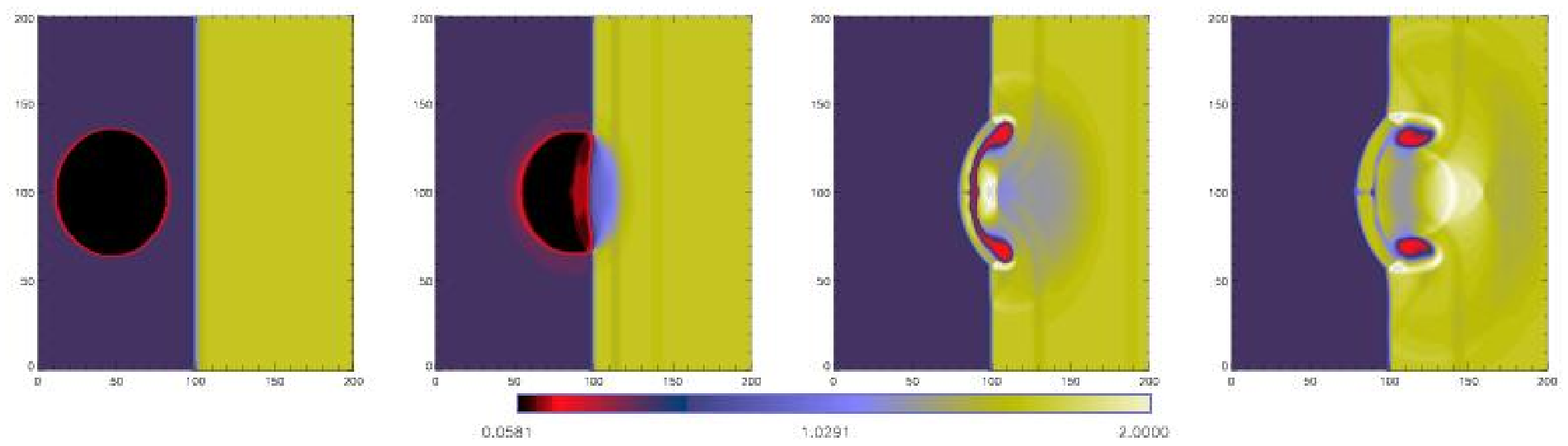}\\
\includegraphics[width=0.49\textwidth]{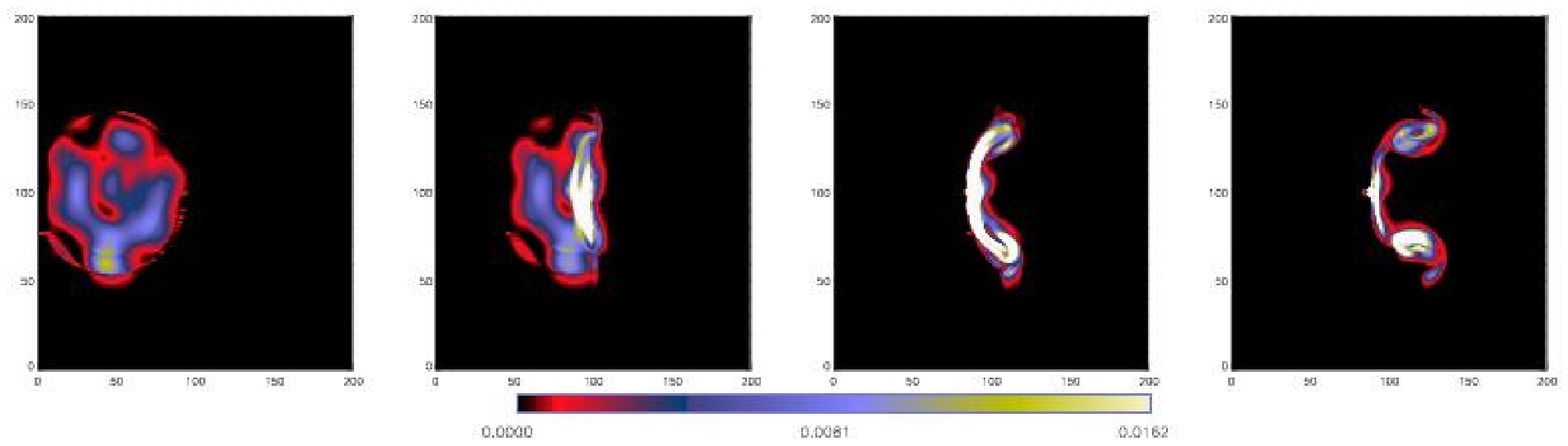}
\caption{{\small Evolution of the
gas density (top) and MF energy density (bottom) for a shock
interacting with fossil radio plasma.  Taken from
\citet{2002MNRAS.331.1011E}.}}
\label{fig:ens_brue_I}
\end{figure}

This idea was investigated  in a more realistic modelling by
\citet*{2004MNRAS.347..389H}, using an SPH {\it Gadget} (\citealt*{SP01.1})
simulation of a merging galaxy cluster within a cosmological
environment.  Such a simulation showed that the probability for a
shock wave to flare the radio plasma is highly suppressed in the
central regions of galaxy clusters compared to the peripheral ones,
where they found illuminated structures up to Mpc in size (see
Fig. \ref{fig:hoeft_I}). The reason for this is that first at the
center the radio plasma ages much faster due to its higher pressure
(and the losses coming in together with the higher MF), and second that the
compression ratio of the shock wave is much higher in the low-density
peripheral regions. It is worth to notice that a necessary condition
to form such relics is that the initial state of
the fossil radio plasma is characterized by a ratio of $P_{\mathrm
B}/P_{\mathrm gas}$ which is as low as 1 per cent to allow shocks to
revive $\approx$ 1-Gyr old radio ghosts. It is also important to
mention that \citet{2004MNRAS.347..389H} find high probability of radio emission outside of
the shocks, related to drained gas flows, induced by the
merger events, which transport material from the outskirts towards the
higher density regions. Thereby in some cases the adiabatic
compression seems to be enough to revive the fossil radio plasma.

\section{Shocks in cosmological simulations}

Cosmological shocks, mainly the accretion shocks on cosmological
objects like galaxy clusters and filaments, are much more frequent than
the ones produced by individual merger events. They also can
produce MFs by the so-called Biermann battery effect
(\citealt{Kulsrud..ApJ.1997,Ryu..1998}) on which a subsequent turbulent
may operate (see section \ref{sec:origin}). In such a
scenario, the MF is strongly correlated with the large-scale structure
(see Fig. \ref{fig:Sigl_I}). This means that in such a case, the MF
within the filamentary structure could be even slightly higher than
its equipartition value without violating the (weak) upper limits of
rotation measure of quasars, as pointed out by \citet{Ryu..1998}.
It is worth to point out that the arguments
for the turbulent dynamo action, which could amplify the battery
seed fields up to $\mu$G level, as presented in
\citet{Kulsrud..ApJ.1997}, refer explicitly to regions about to
collapse into galaxies. It has still to be proven that such
arguments hold within proto-clusters or even cosmological
structures, like sheets and filaments. In general, the time
evolution of the MF as  predicted by these simulations saturates
around  $z\approx3$ (\citealt{Kulsrud..ApJ.1997}) and
leads to a relatively uniform MF strength on scales of tens of
Mpc within the LSS around galaxy clusters (see Fig.
\ref{fig:Sigl_I}). Note that so far there is no comparison of
synthetic rotation measurements obtained by the MFs predicted by
up-scaling the battery fields, with observations on scales of galaxy
clusters. This might be partially motivated by the lack of
resolution in simulated clusters.

\begin{figure}
\includegraphics[width=0.49\textwidth]{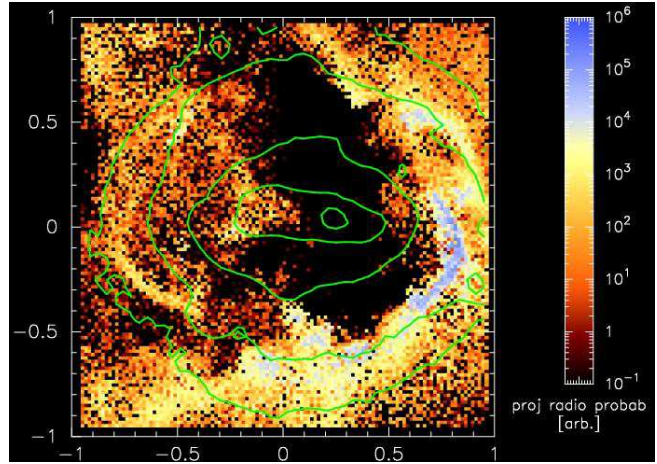}
\caption{{\small The projected
'potential' radio luminosities for 1.13~Gyr old radio plasma, where
$P_B/P_{\rm{gas}} = 0.01$. For comparison the bolometric surface X-ray
luminosity is given. The total bolometric X-ray luminosity of the
cluster is $2\times10^{44}\,\rm{erg}\,\rm{s}^{-1}$ and the
emission-weighted temperature is $3\,\rm{keV}$. Taken from
\citet{2004MNRAS.347..389H}.}} \label{fig:hoeft_I}
\end{figure}

In addition these shocks act as place for acceleration of cosmic
rays (CR) which then will be accreted into the LSS, specially
within galaxy clusters. Using the {\it COSMOR}
(\citealt{2001CoPhC.141...17M}) code, \citet{2001ApJ...562..233M}
followed primary ions and electrons (injected and accelerated by
diffuse cosmic shocks) and secondary electrons and positrons (produced
in p-p inelastic collisions of CR ions with thermal ICM nuclei) within
a cosmological simulation.  Under the assumption that the MF produced
by the battery effect reflects a fair representation of the true
distribution of relative MF strengths within the LSS, they were able
to predict the central radio emission (radio halos) - mainly produced
by secondary CRs - as well as the peripheral radio emission (radio
relics) - mainly produced by primary CRs - in a self-consistent
treatment (see Fig. \ref{fig:Min_I}): the resulting morphology,
polarization and spectral index match the observed properties.
However, one has to notice that the extrapolation of these simulations
(on group scale) to the observed data (i.e. on massive cluster scale)
might not be straightforward (see Fig.
\ref{fig:pt}). It is further worth to notice that even with a
proton injection ratio $R_{e/p}$ of $10^{-2}$ (which is derived
from observations, see Tab.5 in \citealt{2001ApJ...562..233M}), the
predicted luminosities for the primary emission is still significantly
higher than for the secondary one (see Fig. \ref{fig:pt}), which seems
not to be confirmed by the observations. Although the results for the
secondary model are in agreement with previous work on massive galaxy
clusters (\citealt{2000A&A...362..151D}), we notice that generally in the
secondary model radio haloes are predicted for every cluster, which is
in contradiction to the observations (see Fig. \ref{fig:pt}). 

\begin{figure}
\includegraphics[width=0.49\textwidth]{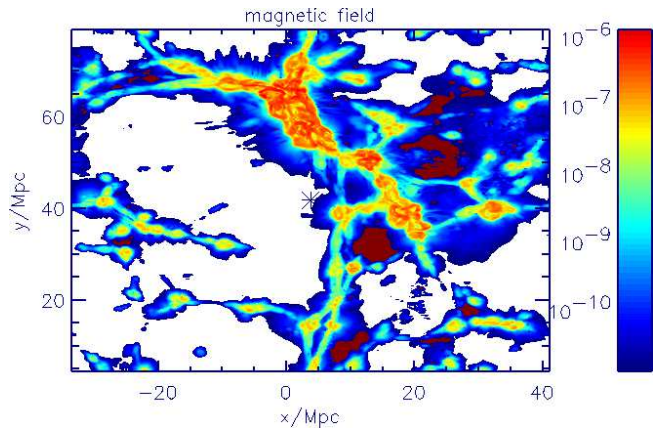}
\caption{{\small A two-dimensional
cut through a cosmological box simulated by
\citet{2001ApJ...562..233M}, and following the evolution of battery
fields. Shown is (the logarithm of) the up-scaled MF strength in
Gauss taken at $z=0$. Taken from \citet*{2004PhRvD..70d3007S}.}}
\label{fig:Sigl_I}
\end{figure}

\section{Cosmological MHD simulations}

Using {\it GrapeMSPH} (\citealt*{1999A&A...348..351D}) and assuming that a
small initial magnetic seed field exists before structure formation
(see section \ref{sec:origin}), the first self-consistent simulations
which follow the MF amplification during the formation of galaxy
clusters with cosmological environment have been performed
(\citealt{1999A&A...348..351D},\citealt*{Dolag:2002}). 
These runs were able to demonstrate that the contribution to the
amplification of MFs by shear flows (and by its induced turbulence) is
significant (see Fig. \ref{fig:b_ampli}). Therefore for the first time
a consistent picture of the MF in galaxy clusters could be
constructed: the amplification predicted by the simulations was
capable to link the predicted strength of the seed MFs (see section
\ref{sec:origin} and references therein) at high redshift ($z\approx3$
and higher) to the today observed MF strength in galaxy clusters.
Furthermore the simulations predicted that the final structure of the
MF in galaxy cluster reflects the process of structure formation, and
no memory on the initial MF configuration survives: this relaxes the
constraints on models for seed MFs. In general such models predict a
MF profile similar to the density profile. Thereby the predicted
rotation measure (RM) profile agrees with the observed one (see Fig.
\ref{fig:RMprof}). \citet{Dolag:2001} found  a quasi linear
correlation between two observables, namely the X-ray surface
brightness and the RM r.m.s.. This result is now confirmed to hold
over several orders of magnitude in a collection of observational data
existing in the literature (see Fig \ref{fig:Lx_RM}).  Extending
{\it Gadget2} to follow the full set of ideal MHD equations,
\citet{2004JETPL..79..583D,2005JCAP...01..009D} performed several
realizations of a cosmological volume confirming the previous findings
even at much higher resolution. Recently,
\citet{2005ApJ...631L..21B} performed a simulation of the
formation of a single galaxy cluster in a cosmological framework,
using a passive MHD solver implemented into FLASH.  Thereby they
nicely confirmed all previous results based on the SPH codes, using
their adaptive mesh refinement (AMR) simulation code (see Fig. \ref{fig:Brueggen_I}).

\begin{figure}
\includegraphics[width=0.49\textwidth]{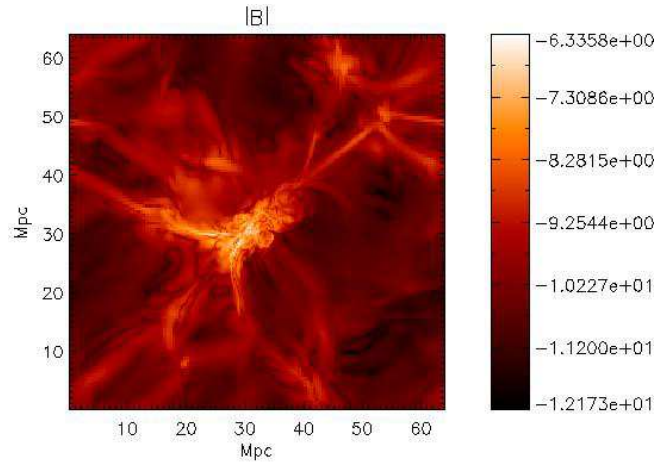}
\caption{{\small Slice through the centre of the simulated box, 
hosting a massive galaxy cluster. The simulation, made using the AMR
code {\it FLASH}, follows the evolution of a weak magnetic seed
field. Shown is the logarithm of the MF strength in Gauss, as
measured at $z=0.5$.  Taken from \citet{2005ApJ...631L..21B}.}}
\label{fig:Brueggen_I}
\end{figure}

Another interesting quantity to look at is the slope $\alpha$ of
the MF power spectrum ($\propto k^{-\alpha}$, with $k$ 
being the wave vector). Within galaxy
clusters $\alpha$ is predicted by the SPH simulations
(\citealt{Dolag:2002},\citealt*{Rordorf:2004}) to be slightly lower, but still
very close to $11/3$, which is the expected value for a
Kolmogorov-like spectrum (in 3D). The AMR simulation by
\citet{2005ApJ...631L..21B} nearly perfectly matches the
Kolmogorov slope. Fig. \ref{fig:T_alpha} shows a collection of
predicted and observed slopes for the MF spectrum. Note that this can
only be a crude comparison, due to the limitations in both
observations and simulations. But in general the
range of slopes predicted by the simulations seems to be consistent
with the slopes inferred from observations of real clusters. It is
worth to notice that the numerical results suggest that the spread of
the slopes might reflect the dynamical stage of the system, 
indicated by the inlay, showing the two mass accretion histories. Here
the smoothly accreting clusters show the
smallest slope (indicating the lack of large-scale power). On the
contrary, the clusters having a quite violent history, with several
major mergers indicated by the small arrows, show a very steep
spectra. Two of them happened in the last 3 Gyrs, thereby inducing an excess
of power on large scales.

Further support for strong MF amplification during the process of
structure formation comes from the application of the Zel'dovich
approximation to follow the MHD equations during the gravitational
collapse (\citealt*{2003MNRAS.338..785B,2005astro.ph..8370K}). Such works
showed super-adiabatic amplification due to the
anisotropy of the collapse of the LSS within the
cold dark matter paradigm.

A novel aspect of including MF pressure into
LSS simulation - even if in a simplified way - is
investigated in an ongoing project (see the contribution by Gazzola), 
which aims at identifying the point for which such non thermal
pressure support starts to significantly modify the structure
formation.

\begin{figure}
\includegraphics[width=0.49\textwidth]{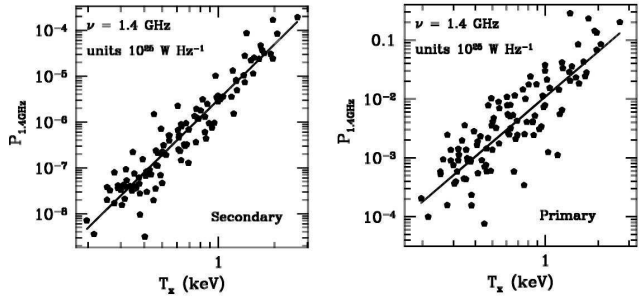}
\caption{{\small Synchrotron
power at 1.4 GHz from secondary electrons (left panel) and primary
electrons (right panel). Note that the values for the luminosities for
primary electrons should be scaled with the electron to proton
injection ratio $R_{e/p}$.  Taken from \citet{2001ApJ...562..233M}. }}
\label{fig:Min_I}
\end{figure}

\section{Conclusive remarks}\label{sec:discussion}

It seems that within the last years a probably consistent picture
of MFs arises from numerical works and observations.  Supported by
simulations of individual events/environments like shear flows,
shock/bubble interactions or turbulence/merging events, a
super-adiabatic amplification of MF is predicted.  It is worth to
notice that this common finding is obtained by using a variety of
different codes, based on different numerical schemes.  Further
support for such super-adiabatic amplifications comes from analytical
estimates of anisotropic collapse making use of the Zel'dovich
approximation.  When applied in fully consistent cosmological
simulations, various observational aspects are reproduced (see
Figs. \ref{fig:RMprof} and \ref{fig:Lx_RM}) and the resulting MF
amplification reaches a level sufficient to link models
predicting seed MFs at high redshift with the MFs observed in galaxy
clusters today. It is important to mention that all simulations show
that this effect increases when the resolution is improved, and
therefore all the numbers have to be taken as lower limits of possible
amplification.

\begin{figure}
\includegraphics[width=0.49\textwidth]{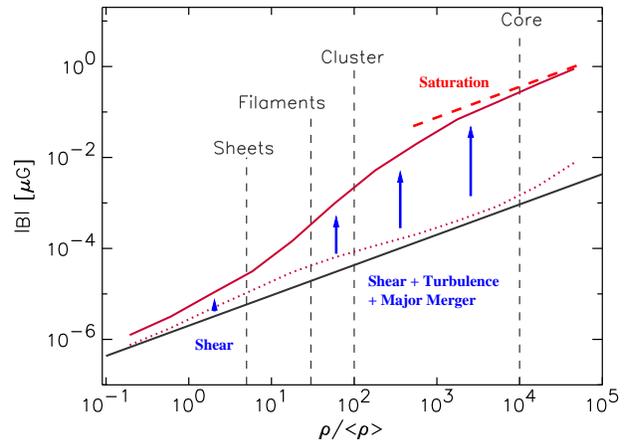}
\caption{{\small The MF strength as a function of baryonic overdensity.
The long-dashed line shows the expectation for a purely adiabatic
evolution, the solid line gives the mean field strength at a given
overdensity within a cosmological simulation
(\citealt{2005JCAP...01..009D}). While the
latter is close to the adiabatic value in underdense regions, there is
a significant inductive amplification in clusters due to shear flows
and turbulence, subject however to saturation in the cluster cores. At
any given density, a large fraction of particles remains close to the
adiabatic expectation, as shown by the dotted line, which gives the
median of the distribution at each density.}}
\label{fig:b_ampli}
\end{figure}

Note that on the other hand there are significant differences for the
predictions of the MF structures coming from different models of
seed MFs. In particular there are main differences
between the up-scaled, cosmological battery fields
(\citealt{2001ApJ...562..233M,2004PhRvD..70d3007S}) and the MF
predicted from high-resolution simulations of galaxy clusters
using either AMR (\citealt{2005ApJ...631L..21B}) or SPH
(\citealt{1999A&A...348..351D,Dolag:2002,2005JCAP...01..009D}). 
In the latter case it is possible to follow the amplification of seed
fields within the turbulent ICM in more detail. A good visual
impression can be obtained by comparing the regions filled with high
MFs shown in Figs. \ref{fig:Sigl_I} and \ref{fig:Brueggen_I}. It is
clear that the high MF regions for the battery fields are predicted to
be much more extended, leading to a flat profile around the forming
structure, whereas for the turbulent amplified MFs the clusters show a
much more peaked MF distribution. Part of this difference
originates from the physical model behind, as the cosmological shocks
are much stronger outside the clusters than inside.  Somewhat more
unclear is what the contribution of the different numerical resolutions
of these simulations to these discrepancies is. Calibrating such
simulations using the MF measured only in the high-density regions of
galaxy clusters makes it crucial to perform a more detailed comparison
with all available observations. Note that extrapolating the
predictions of the simulations into lower density regions, where no
strong observational constraints exist, will even amplify the
differences in the predictions for the MF structure between the
different simulations.

Moreover, one has to keep in mind that depending on the ICM
resistivity the MF could be suffering decay, which so far is
neglected in all the simulations discussed before.  Furthermore a
clear lack of the present simulations is that they do not include the
creation of MF by all the feedback processes happening within LSS
(like radio bubbles inflated by AGNs, galactic winds, etc.): this
might alter the MF prediction if their contribution turns out to be
significant. Also all the simulations done so far neglect radiative
losses: if included, this would lead to a significant increase of the
density in the central part of clusters and thereby to a further MF
amplification in these regions.  Finally, there is an increasing number
of arguments suggesting that instabilities and turbulence on very
small scales can amplify the MFs in a relatively short timescale
reaching the observed $\mu$G level. However it is still unclear how
such very small scale fields can be ordered on large (up to hundreds
of kpc) scales as observed.

\begin{figure}[t]
\includegraphics[width=0.49\textwidth]{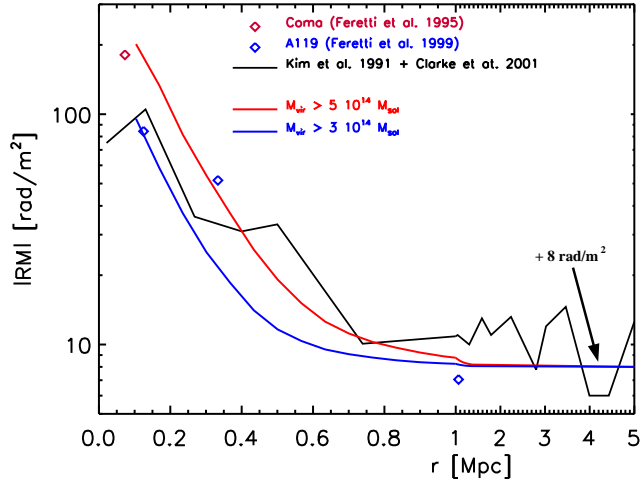}
\caption{{\small Comparison of RMs from the simulation with observations
for Abell clusters, as a function of distance to the closest
cluster. Smooth lines represent the median values of $|{\rm RM}|$
produced by simulated clusters with masses above $5 \times
10^{14}~{\rm M}_\odot$ and $3 \times 10^{14}~{\rm M}_\odot$. The
broken line represents the median of combined data taken from the
independent samples presented in \citet{1991ApJ...379...80K} and
\citet{2001ApJ...547L.111C}. We also include data (diamonds) for
the three elongated sources observed in A119
(\citealt{1999A&A...344..472F}), and for the elongated source observed
in the Coma cluster (\citealt{1995A&A...302..680F}).}}
\label{fig:RMprof}
\end{figure}

Concerning Radio haloes and relics the picture is only partially
consistent ( for a more detailed discussion of primary and
secondary models see \citealt{2004JKAS...37..493B} and
references therein).
Radio relics seem to be most likely related to strong
shocks produced by major merging events and therefore produced by
direct re-acceleration of CRs, the so-called primary models. Although
some of the observed features like morphology, polarization and
position with respect to the cluster centre can be resonably well reproduced,
there might be still some puzzles to solve. On one hand, direct
acceleration of CRs in shocks seems to overestimate the abundance and
maybe the luminosity of radio relics; on the other hand simulations
which illuminate fossil radio plasma can produce reasonable relics
only starting from a small range of parameter settings.  A similar
situation arises for modelling the central radio emission of galaxy
clusters by secondary models. On one hand, the total luminosity seems
to be reproduced using reasonable assumptions and also the steep
observed correlation between cluster temperature/mass and radio power
seems to be reproduced quite well. But such models suffer from two
drawbacks. The first one is that in the framework of such models every
massive cluster produces a powerful radio halo, but this is not
confirmed by observations (see Fig. \ref{fig:pt}).  The other problems
is that the detailed radio properties are not reproduced. Firstly, the
profile of radio emission in most cases is too steep, so that these
models cannot reproduce the size of the observed radio halos in almost
all cases. Secondly, the observed spectral steepening (e.g. 
\citealt{2005A&A...440..867G}) also cannot be reproduced. Note that there
is also no indication from observations that clusters showing radio
emission contain higher MFs than the ones without observable extended,
diffuse radio
emission. On the contrary, the cluster A2142 has a MF strength similar
to the Coma cluster (see Fig. \ref{fig:Lx_RM}), but the upper limit on
its radio emission is at least two orders of magnitude below the value
expected from the correlation (see Fig. \ref{fig:pt}).  Note that both
clusters are merging systems characterized by the presence of two
central cD galaxies. This indicates that there should be further
processes involved or additional conditions to fulfill to produce
radio emission. It is worth to notice that recent models, based on
turbulent acceleration, seem to overcome this problem
(see \citealt*{2003ApJ...594..732K,2005MNRAS.363.1173B,2005MNRAS.357.1313C} and references
therein).
Note that \citet{2005MNRAS.357.1313C}
predicted the probability for a galaxy cluster to show giant radio halo
to be an increasing function of cluster mass and reproduced the observed
fraction of $\approx 30$\% for massive galaxy clusters.

There is a big challenge for the next generation of cosmological MHD
simulations. Simulations are quite close to having the resolution
necessary to properly describe the MF components down to
the observed scales, but on the other hand this means to resolve
galaxies inside the simulations as well.  Note that following the
dynamics of such structures for which the MF might
dominate the evolution, is a real challenge within LSS simulations.
However, we expect that - if succeeding in overcoming these
limitations - the dynamical impact of the MF on regions
like the cooling flows at the centres of galaxy clusters will be
significant and will contribute to solve these outstanding puzzles.

\begin{figure}[t]
\includegraphics[width=0.49\textwidth]{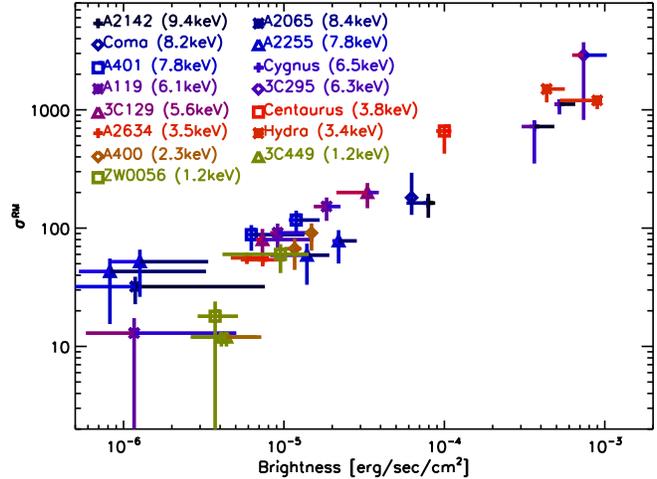}
\caption{{\small Correlation of rotation measurements with X-ray surface
brightness for different galaxy clusters. The data points are
observations collected from the literature. The colors represent
the virial cluster temperature taken from literature. This is an
updated version of the correlation presented in
\citet{Dolag:2001}.}} \label{fig:Lx_RM}
\end{figure}

\acknowledgements This work 
%has been carried out under the HPC-EUROPA
%project (RII3-CT-2003-506079), with the support of the European
%Community - Research Infrastructure Action under the FP6 ``Structuring
%the European Research Area'' Programme, and under 
%RadioNet R113CT 2003 5058187.
has benefitd from research funding from the European Community's
sixth Framework Programme under the HPC-EUROPA
project (RII3-CT-2003-506079) and under RadioNet (R113-CT-2003-5058187).
I want to thank the INAF-IRA in
Bologna and the Astronomy Department in Trieste for their
hospitality. Many thanks to S. Borgani and L. Moscardini for
stimulating discussions and comments. Special thanks also to F. Govoni 
who helped a lot in preparing many of the observation-related figures.

\begin{figure}[t]
\includegraphics[width=0.49\textwidth]{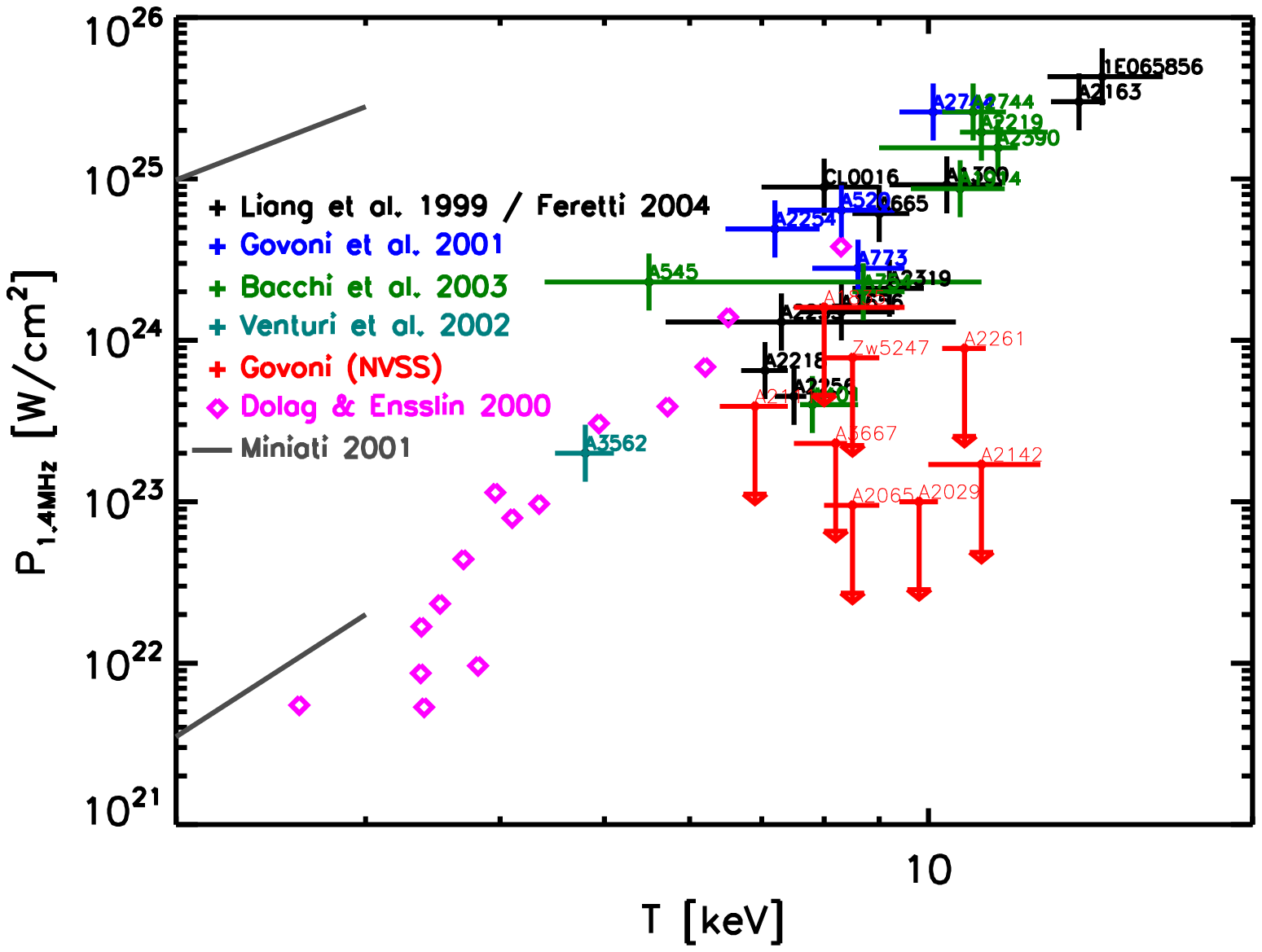}
\caption{{\small Total power of radio halos observed at 1.4 MHz vs.
cluster temperature. We plot data from
\citet{2000ApJ...544..686L}, which were partially re-observed by
Feretti (2005, in preparation) together with data from
\citet{2001A&A...369..441G,2003A&A...400..465B,2003A&A...402..913V}.
Some additional upper limits are collected with the help of F.
Govoni. We applied a secondary hadronic model as described in
\citet{2000A&A...362..151D} to calculate the radio emission from
the simulated galaxy clusters. We also added the predictions for
the emission from primary (upper) and secondary (lower) electrons
taken from \citet{2001ApJ...562..233M}. Note that the values for
the luminosities for primary electrons should be scaled with the
electron to proton injection ratio $R_{e/p}$.}} \label{fig:pt}
\end{figure}

\begin{figure}
\includegraphics[width=0.49\textwidth]{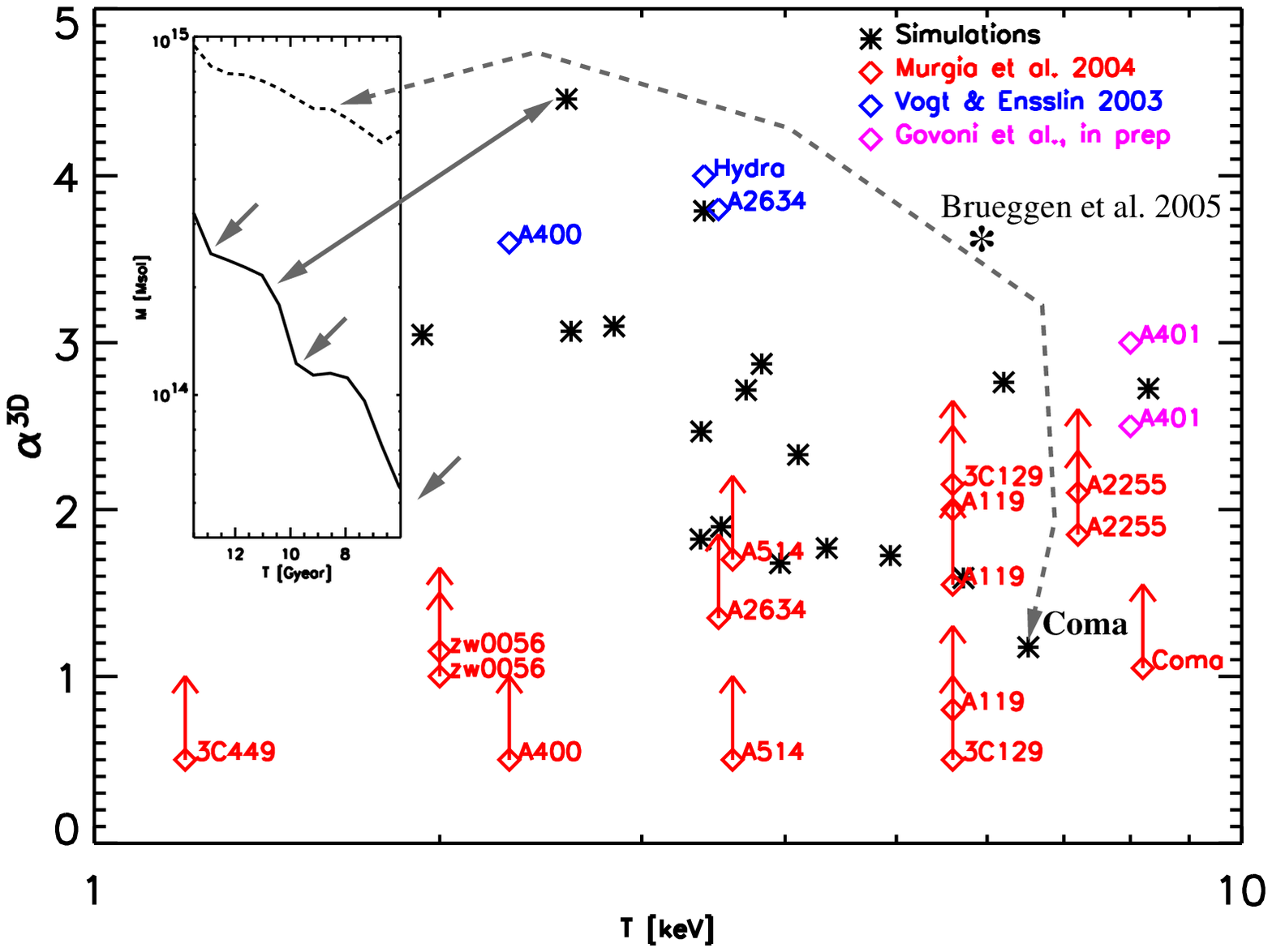}
\caption{{\small Rough comparison of observed index (diamonds) of MF power
spectrum (\citealt{2003A&A...412..373V,2004A&A...424..429M}) and the
predictions (stars) from simulations
(\citealt{Rordorf:2004,2005JCAP...01..009D,2005ApJ...631L..21B}). Note
that some observations reflect only a weak lower limit.  Morever the
displayed data spawn different environments, length scales and
techniques used to constrain the spectral index. Furthermore the
spectral index calculated from the simulations may suffer from effects
caused by a non constant mean MF strength in the cluster and varying
numerical resolution.}} \label{fig:T_alpha}
\end{figure}

\bibliographystyle{mn2e}
\bibliography{master2}

\end{document}